\documentclass[
reprint,
aip, jap,%cha,%
superscriptaddress,
floatfix%,linenumbers
]{revtex4-1}

\usepackage{makeidx}
\usepackage[verbose]{placeins}
\usepackage{multirow}
\usepackage{graphicx}
\usepackage{amsmath}
\usepackage{epstopdf}
\usepackage{textcomp}
\usepackage{epstopdf}
\usepackage{tabularx}
\usepackage{ulem} % allow the strikethrough command \sout
\usepackage{color}

\usepackage{natbib}
\usepackage{bm}%
\usepackage[colorlinks=true,linkcolor=blue]{hyperref}%
\expandafter\ifx\csname package@font\endcsname\relax\else
 \expandafter\expandafter
 \expandafter\usepackage
 \expandafter\expandafter
 \expandafter{\csname package@font\endcsname}%
\fi
\makeindex

\begin{document}

\title{Relative weight of the inverse spin Hall and spin rectification effects for metallic Py,Fe/Pt and insulating YIG/Pt bilayers estimated by angular dependent spin pumping measurements}

\author{S. Keller}
\author{J. Greser}
\author{M. R. Schweizer}
\author{A. Conca}
\author{B. Hillebrands}
\author{E. Th. Papaioannou\email{papaio@rhrk.uni-kl.de}}
%\thanks{Author to whom correspondence should be addressed. Email: papaio@rhrk.uni-kl.de}

\affiliation{Fachbereich Physik and Landesforschungszentrum OPTIMAS, Technische Universit\"{a}t Kaiserslautern,
Erwin-Schr\"{o}dinger-Str. 56, 67663 Kaiserslautern, Germany}

\date{\today}

\begin{abstract}

We quantify the relative weight of inverse spin Hall and spin rectification effects occurring in RF-sputtered polycrystalline permalloy, molecular beam epitaxy-grown epitaxial iron and liquid phase epitaxy-grown yttrium-iron-garnet bilayer systems with different capping materials. To distinguish the spin rectification signal from the inverse spin Hall voltage the external magnetic field is rotated in-plane to take advantage of the different angular dependencies of the prevailing effects. We prove that in permalloy anisotropic magnetoresistance is the dominant source for spin rectification while in epitaxial iron the anomalous Hall effect has an also comparable strength. The rectification in yttrium-iron-garnet/platinum bilayers reveals an angular dependence imitating the one seen for anisotropic magnetoresistance caused by spin Hall magnetoresistance.

\end{abstract}



\maketitle {}

%\section{Introduction}

Spintronic bilayers composed of a ferromagnetic (FM) and a nonmagnetic (NM) layer with large spin-orbit-interaction are promising devices for the spin-to-charge conversion for future applications. At ferromagnetic resonance (FMR) the spin pumping (SP) effect allows for the injection of a pure spin current from the FM into the NM layer \cite{Bauer2002}. There, the spin current is converted into a charge current by the inverse spin Hall effect (ISHE) \cite{Saitoh2006}. A wide range of metallic, semiconducting or insulating ferro- \cite{Andres} and ferrimagnets \cite{goennewein2011} and NM \cite{azevedo} materials, like Au, Pd, Ta, W, and Pt, have been investigated up to this point. In metallic FM layers, an overlapping additional effect take place, the so called spin rectification (SR) effect, which hinders the access to the pure ISHE signal.
Different approaches for separation have been thoroughly investigated \cite{Evangelos, Saitoh2016, Harder2016}. Thickness variations of the FM and the NM layers show different dependencies for ISHE and SR, but require a lot of effort for producing whole sample series or wedged microstructures. Another method is a sweep of the excitation frequency, which cannot be applied to all experimental setups and requires a careful calibration of the microwave transmission properties of the setup. Also the minimization of the electrical microwave field at the location of the sample by using a microwave cavity is possible, but in most cases only fixed frequencies can be applied.
The rotation of the magnetization angle by rotating the external magnetic field in- or out-of-plane is one of the most common and practicable methods, with out-of-plane rotation normally requiring larger magnetic fields for thin films \cite{Saitoh2016, Harder2016}. Here, we quantify with the help of the in-plane angular dependent spin pumping measurements the ISHE and the SR contributions, mainly anisotropic magnetoresistance (AMR) and anomalous Hall effect (AHE). Therefore, bilayers composed of magnetic (Fe, Py, and YIG) and non-magnetic (Pt, Al and MgO) materials have been used. Capping layers with significant spin Hall angle $\Theta_{\mathrm{SH}}$ (Pt) should show a large ISHE, while materials with small $\Theta_{\mathrm{SH}}$ (Al) and insulating materials (MgO) should not. All bilayer samples with metallic FM (Py/Al, Py/Pt, Fe/MgO and Fe/Pt) have the dimensions of $(10 \times 10)$ mm$^2$, while the YIG/Pt sample is of smaller dimensions $(2 \times 3)$ mm$^2$.

%\section{Probing of the spin pumping by means of VNA-FMR spectroscopy}

\begin{figure}
\includegraphics[width =0.95 \columnwidth]{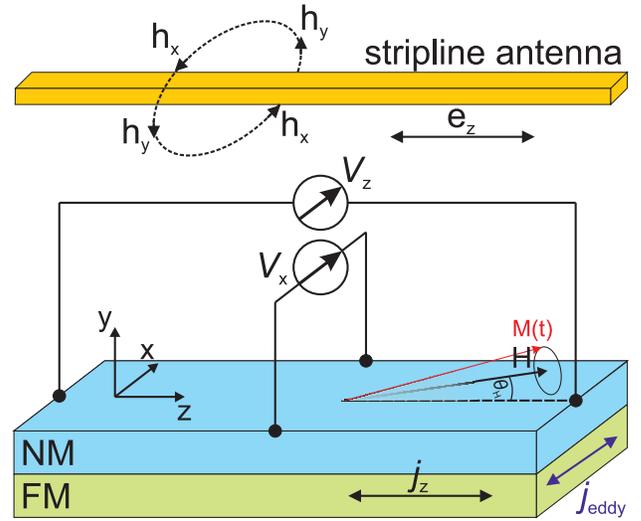}
\caption{\label{fig:SRKS} Experimental setup and coordinate system: $x$, $y$ and $z$ are the lab fixed coordinates. The external field $\vec{H}$ rotates in-plane, while the angle $\Theta_{\mathrm{H}}$ is defined as the angle between $z$ and $\vec{H}$. The bilayer films are lying in the $x$ and $z$ plane and $y$ is the out-of-plane coordinate. The exciting stripline antenna is parallel to $z$ and is generating an in-plane dynamic magnetic field $h_{x}$, an out-of-plane field $h_y$ and also a dynamic electrical field $e_{z}$, which induces an electrical current $j_{z}$ in the samples in $z$ direction. Eddy currents $j_{\mathrm{eddy}}$ potentially can flow transverse to the microwave electrical field in $x$ direction. The electrical contacts for measuring the DC voltage  are either transverse ($V_{x}$) or parallel ($V_{z}$) to the stripline antenna.}
\end{figure}

\begin{figure}
\includegraphics[width =0.95 \columnwidth]{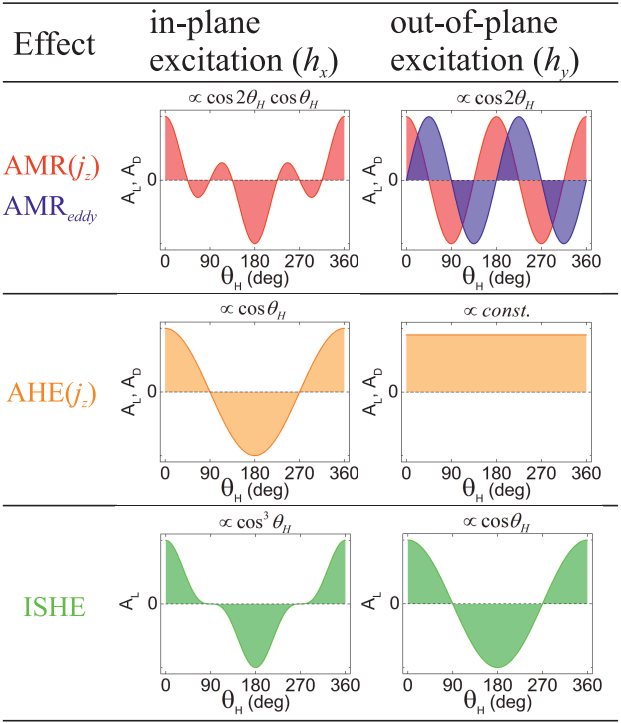}
\caption{\label{fig:SR5} Theoretical in-plane magnetization angular dependencies of spin rectification effects and ISHE with contacts transverse to the microwave antenna ($x$ direction) and different dynamic magnetic field geometries, adapted from Harder et.~al.\cite{Harder2016}. $\Theta_{\mathrm{H}}$ is the magnetic field angle (defined in Fig.~\ref{fig:SRKS}), $A_{\text{L}}$ and $A_{\text{D}}$ are the amplitudes of the effects contributing to the symmetric voltage (L: Lorentzian) and the antisymmetric voltage (D: Dispersive).}
\end{figure}

We first address the measurements on polycrystalline Py/Pt and Py/Al bilayers, that is, with presence and absence of ISHE voltage, respectively. We will use the data of this model system to illustrate the angular dependence of the measured signal and the analysis method used to separate the different contributions. Second, we will present the data for epitaxial Fe/Pt and Fe/MgO samples and we will apply again the same analysis method comparing the weights of the different contributions with the Py case. Finally, results in YIG/Pt bilayers are presented to compare the situation for a system with an insulating magnetic layer where no AMR or AHE can be present. Prior to concluding, we will present some important remarks about the validity and limitations of the analysis method based on angular measurements.

In the experiment for a fixed excitation frequency and external field angle, the external field amplitude is swept. The voltage measured by lock-in-amplification technique exhibits peaks consisting of symmetric and antisymmetric components which are fitted by the following equation for each individual external magnetic field sweep \cite{azevedo}:

\begin{equation}\label{SPfit}
\begin{split}
V_{\text{meas}}(H) = &  V_{\text{sym}} \frac {(\Delta H)^{2} }{(H-H_{\text{FMR}}) ^{2} + {(\Delta H)^{2} }}\\ 
& + V_{\text{asym}} \frac { -2\Delta H(H-H_{\text{FMR}}) } {(H-H_{\text{FMR}}) ^{2} + (\Delta H)^{2} },
\end{split}
\end{equation}

\noindent where $V_{\mathrm{sym}}$ and $V_{\mathrm{asym}}$ are the amplitude of the  symmetric and antisymmetric components, respectively. $\Delta H$ is the linewidth, $H$ is the applied magnetic field, and $H_{\mathrm{FMR}}$ is the corresponding FMR field value. While the SP/ISHE effect contributes only to  $V_{\mathrm{sym}}$, the SR effects contribute to both voltage amplitudes. The relative contribution of AMR to $V_{\mathrm{sym}}$ and $V_{\mathrm{asym}}$ and AHE to $V_{\mathrm{sym}}$ and $V_{\mathrm{asym}}$ is determined by the phase difference between the dynamic magnetization $\vec{m}(t)$ and the microwave electrical field induced AC current $\vec{j}(t)$ inside the FM layer. This phase difference is not easily accessible \cite{azevedo} and the relative contribution of AMR does not necessarily have to be the same as the one of AHE \cite{Harder2016}. To fit the measured voltage amplitudes it is needed to calculate the angular dependencies of SP/ISHE and SR (a detailed derivation can be found in \cite{Harder2016, Saitoh2016}). The symmetric as well as the antisymmetric voltage will then be fitted. 

\begin{figure}
\includegraphics[width =0.95 \columnwidth]{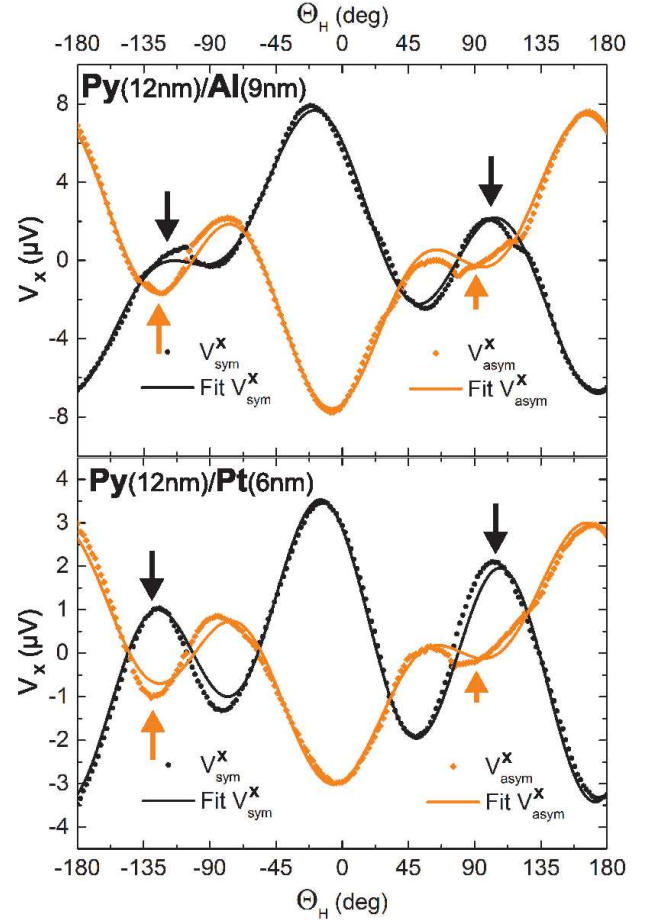}
\caption{\label{fig:Py} Angular dependent spin pumping measurements of Py(12nm)/Al(10nm) (top graph) and Py(12nm)/Pt(10nm) (bottom graph) at 13 GHz excitation frequency with contacts transverse to the direction of the stripline antenna. Black and orange arrows are highlighting the side-maxima/minima originating from AMR.}
\end{figure}

First let us consider the coordinate system (see Fig.~\ref{fig:SRKS}), where $x$ and $z$ are the in-plane and $y$ the out-of-plane lab fixed coordinates, $\Theta_\mathrm{H}$ is the angle between the external magnetic field $\vec{H}$ and the $z$ axis. The electrical contacts are either in $x$ (transverse to the stripline antenna) or $z$ (parallel to the stripline antenna) direction. $j_{z}$ induced by the microwave electrical field $e_{z}$ and $j_{\mathrm{eddy}}$ in $x$ direction (explained later) are the in-plane current components. The dynamic magnetic microwave fields $h_{x}$ (in-plane) and $h_{y}$ (out-of-plane) are determined by the microwave stripline antenna.

\begin{table*}
	\centering
		\begin{tabular} {cccrclcrclrclcrclcrcl}
		$V_{\mathrm{sym}}/V_{\mathrm{asym}}$	&Sample & $V^{h_x}_{\mathrm{ISHE}}$ ($\mu V$)&\multicolumn{3}{c}{$V^{h_y,h_x}_{\mathrm{ISHE,AHE}}$ $^\dagger$ ($\mu V$)} & ~ &\multicolumn{3}{c}{$V^{h_y}_{\mathrm{AHE}}$ ($\mu V$)}  & \multicolumn{3}{c}{$V^{h_x,j_z}_{\mathrm{AMR}}$ ($\mu V$)} & ~ & \multicolumn{3}{c}{$V^{h_y,j_\mathrm{z}}_{\mathrm{AMR}}$ ($\mu V$)} & ~ & \multicolumn{3}{c}{$V^{h_y,j_\mathrm{eddy}}_{\mathrm{AMR}}$ ($\mu V$)}\\
			\hline
\multirow{4}{*}{$V_{\mathrm{sym}}$}	&	Py/Al & 0  & {\bf 1.43} &\color[rgb]{0.5,0.5,0.5}$\pm$&\color[rgb]{0.5,0.5,0.5}0.04 & ~ & 0.15 &\color[rgb]{0.5,0.5,0.5}$\pm$&\color[rgb]{0.5,0.5,0.5}0.02  & {\bf 5.63}&\color[rgb]{0.5,0.5,0.5}$\pm$&\color[rgb]{0.5,0.5,0.5}0.06 & ~ & \multicolumn{3}{c}{0} & ~ & {\bf -1.68}&\color[rgb]{0.5,0.5,0.5}$\pm$&\color[rgb]{0.5,0.5,0.5}0.03 \\
											&	Py/Pt & amb. & \multicolumn{3}{c}{amb.} & ~ & 0.02 &\color[rgb]{0.5,0.5,0.5}$\pm$&\color[rgb]{0.5,0.5,0.5}0.02  & \multicolumn{3}{c}{amb.} & ~ & \multicolumn{3}{c}{0} & ~ &{\bf -0.61}&\color[rgb]{0.5,0.5,0.5}$\pm$&\color[rgb]{0.5,0.5,0.5}0.02 \\
											&	Fe/MgO & 0 & {\bf 6.85}&\color[rgb]{0.5,0.5,0.5}$\pm$&\color[rgb]{0.5,0.5,0.5}0.12 & ~ & -0.01 &\color[rgb]{0.5,0.5,0.5}$\pm$&\color[rgb]{0.5,0.5,0.5}0.05  & {\bf 5.12}&\color[rgb]{0.5,0.5,0.5}$\pm$&\color[rgb]{0.5,0.5,0.5}0.15 & ~ & 0.18&\color[rgb]{0.5,0.5,0.5}$\pm$&\color[rgb]{0.5,0.5,0.5}0.09 & ~ & \multicolumn{3}{c}{0} \\
											&	Fe/Pt & amb. & \multicolumn{3}{c}{amb.} & ~ & 0.12 &\color[rgb]{0.5,0.5,0.5}$\pm$&\color[rgb]{0.5,0.5,0.5}0.05  & \multicolumn{3}{c}{amb.} & ~ & -0.25&\color[rgb]{0.5,0.5,0.5}$\pm$&\color[rgb]{0.5,0.5,0.5}0.08 & ~ & \multicolumn{3}{c}{0}\\
			\hline
\multirow{4}{*}{$V_{\mathrm{asym}}$}	&	Py/Al & 0 &{\bf -1.49}&\color[rgb]{0.5,0.5,0.5}$\pm$&\color[rgb]{0.5,0.5,0.5}0.05 & ~ & 0.03 &\color[rgb]{0.5,0.5,0.5}$\pm$&\color[rgb]{0.5,0.5,0.5}0.03  & {\bf -5.95}&\color[rgb]{0.5,0.5,0.5}$\pm$&\color[rgb]{0.5,0.5,0.5}0.07 & ~ & \multicolumn{3}{c}{0} & ~ &{\bf 1.02}&\color[rgb]{0.5,0.5,0.5}$\pm$&\color[rgb]{0.5,0.5,0.5}0.04 \\
											&	Py/Pt & 0 &{\bf -0.61}&\color[rgb]{0.5,0.5,0.5}$\pm$&\color[rgb]{0.5,0.5,0.5}0.03 & ~ & 0.01 &\color[rgb]{0.5,0.5,0.5}$\pm$&\color[rgb]{0.5,0.5,0.5}0.01  & {\bf -2.36}&\color[rgb]{0.5,0.5,0.5}$\pm$&\color[rgb]{0.5,0.5,0.5}0.04 & ~ & \multicolumn{3}{c}{0} & ~ &{\bf 0.44}&\color[rgb]{0.5,0.5,0.5}$\pm$&\color[rgb]{0.5,0.5,0.5}0.02 \\
											&	Fe/MgO & 0 & {\bf 4.07}&\color[rgb]{0.5,0.5,0.5}$\pm$&\color[rgb]{0.5,0.5,0.5}0.15 & ~ & 0.07 &\color[rgb]{0.5,0.5,0.5}$\pm$&\color[rgb]{0.5,0.5,0.5}0.06  & {\bf -6.20}&\color[rgb]{0.5,0.5,0.5}$\pm$&\color[rgb]{0.5,0.5,0.5}0.19 & ~ & 0.18&\color[rgb]{0.5,0.5,0.5}$\pm$&\color[rgb]{0.5,0.5,0.5}0.10 & ~ & \multicolumn{3}{c}{0}\\
											&	Fe/Pt & 0 & {\bf 3.55}&\color[rgb]{0.5,0.5,0.5}$\pm$&\color[rgb]{0.5,0.5,0.5}0.09 & ~ & 0.01 &\color[rgb]{0.5,0.5,0.5}$\pm$&\color[rgb]{0.5,0.5,0.5}0.04  & {\bf -7.88}&\color[rgb]{0.5,0.5,0.5}$\pm$&\color[rgb]{0.5,0.5,0.5}0.11 & ~ & 0.13&\color[rgb]{0.5,0.5,0.5}$\pm$&\color[rgb]{0.5,0.5,0.5}0.05 & ~ & \multicolumn{3}{c}{0}\\
											
			\hline
		\end{tabular}
	\caption{Results of the angular spin pumping measurements: symmetric and antisymmetric voltage amplitudes of Py/Al, Py/Pt, Fe/MgO and Fe/Pt. Items marked with amb. are ambiguous (see text). The voltage of the effects mainly contributing are marked in bold. The voltage marked with $^\dagger$ corresponds to the term $\propto \cos(\Theta_\mathrm{H})$, which is comprised of in-plane AHE and out-of-plane ISHE in the symmetrical voltage and only of in-plane AHE in the antisymmetric voltage (to be seen in Fig.~\ref{fig:SRKS}). Absolute values between samples are not comparable because of different excitation frequencies.}
	\label{tab:results}
\end{table*}

At first the model of the measurements, where the DC voltage is measured in $x$ direction (transverse to the antenna, shown in Fig.~\ref{fig:SRKS}), is discussed: For this measurement configuration significant values for $j_{z}$ and $h_{x}$ (in-plane dynamic magnetic field component), and smaller values for $h_{y}$ (out-of-plane dynamic magnetic field component), which is estimated a magnitude smaller than the in-plane field components, are considered. The theoretical angular dependencies of the underlying effects are graphically shown in Fig.~\ref{fig:SR5}\cite{Saitoh2016, Harder2016}. It can be recognized that in-plane excited AHE is similar to in-plane excited ISHE bearing only one maximum and one minimum, but with different slopes at zero crossing. In-plane excited AMR is showing three maxima/minima where one is of higher amplitude (referred to as main maximum/minimum in the following) and two of smaller amplitude (referred to as side maxima/minima). Out-of-plane excited AMR is showing two maxima/minima with equal amplitude. Out-of-plane AHE will generate a constant offset and out-of-plane ISHE has an identical shape as in-plane AHE and can therefore not be distinguished from it. As to be shown later in the measurements for the Py samples, an additional AMR effect also takes place. This AMR effect is shown in Fig.~\ref{fig:SR5} in blue and is the only one antisymmetric around 0$^\circ$. This AMR scales with an electrical current $j_{x}$ perpendicular to the microwave induced currents and with an out-of-plane microwave field component $h_{y}$ and is affiliated to eddy currents~\cite{eddy}. To fit the experimental data all considered effects are linear superimposed:

\begin{equation}
\begin{split}
V^{x}_{\mathrm{sym}} ~=~~ & V^{h_x}_{\mathrm{ISHE}}\cos^3(\Theta_\mathrm{H}) +
								  V^{h_y,h_x}_{\mathrm{ISHE,AHE}}\cos(\Theta_\mathrm{H}) ~+\\
								 & V^{h_y}_{\mathrm{AHE}} +
								  V^{h_x,j_z}_{\mathrm{AMR}}\cos(2\Theta_\mathrm{H})\cos(\Theta_\mathrm{H}) ~+\\
								 & V^{h_y,j_z}_{\mathrm{AMR}}\cos(2\Theta_\mathrm{H}) +
								  V^{h_y,j_\mathrm{eddy}}_{\mathrm{AMR}}\sin(2\Theta_\mathrm{H}).\\ & \\
V^{x}_{\mathrm{asym}} =~~ & V^{h_x}_{\mathrm{AHE}}\cos(\Theta_\mathrm{H}) + 
								  V^{h_y}_{\mathrm{AHE}} ~+\\
								 & V^{h_x,j_z}_{\mathrm{AMR}}\cos(2\Theta_\mathrm{H})\cos(\Theta_\mathrm{H}) ~+\\
								 & V^{h_y,j_z}_{\mathrm{AMR}}\cos(2\Theta_\mathrm{H}) +
								  V^{h_y,j_\mathrm{eddy}}_{\mathrm{AMR}}\sin(2\Theta_\mathrm{H}).
\end{split}
\label{fit1}
\end{equation}

\begin{figure}
\includegraphics[width =0.95 \columnwidth]{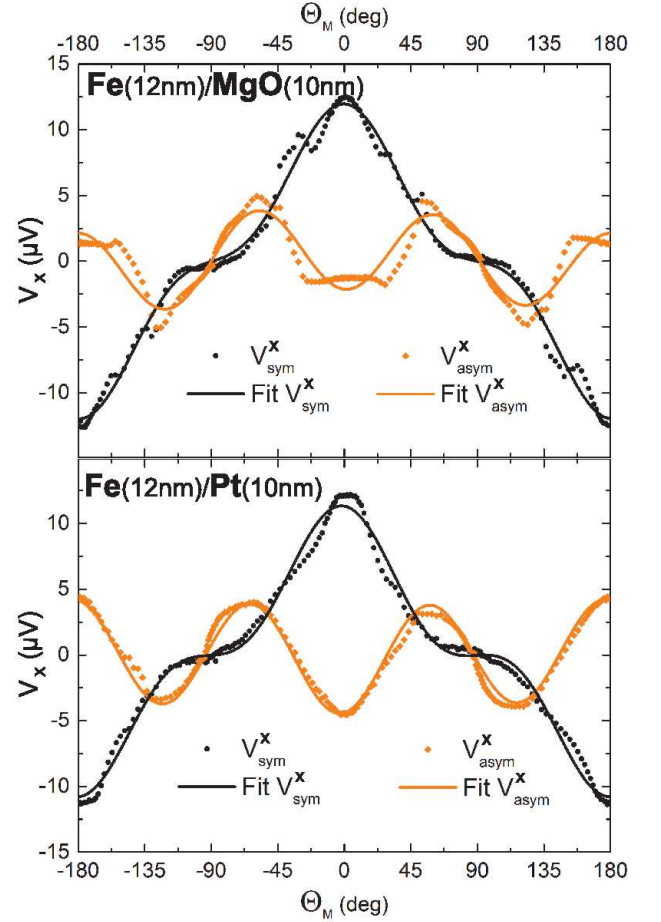}
\caption{\label{fig:Fe} Angular dependent spin pumping measurements of Fe(12nm)/MgO(10nm) (top graph) and Fe(12nm)/Pt(10nm) (bottom graph) at 13 GHz excitation frequency with contacts transverse to the direction of the stripline antenna.}
\end{figure}

Equations~\ref{fit1} were then used to fit the angular dependent spin pumping measurements shown in Fig.~\ref{fig:Py} and~\ref{fig:Fe} for the  of $V_{\mathrm{sym}}$ and $V_{\mathrm{asym}}$ of Py/Al, Py/Pt,  Fe/MgO and Fe/Pt bilayers. The voltage amplitudes from the fits of the Py and Fe sample measurements have been summarized in Table~\ref{tab:results} for comparison. 

After familiarizing with the angular dependencies of the ISHE and SR effects the measurements for the Py bilayers are now discussed: In Fig.~\ref{fig:Py} we see for the Py/Al sample that the signal is mainly consisting of AMR in the symmetric as well as in the antisymmetric amplitude, since the signals exhibit pronounced side-maxima (arrows). The antisymmetric voltage amplitude of Py/Pt has almost identical shape as the one of Py/Al. For both samples the AMR to AHE ratio of the antisymmetric voltage is approximately 1 to 4 (see Table~\ref{tab:results}). Py/Al and Py/Pt also show that their side-maxima (arrows) are having not the same amplitudes. This is correlated to AMR caused by eddy currents with an out-of-plane dynamic magnetic field component (see Table~\ref{tab:results}).

The measurements of Fe/MgO and Fe/Pt can be seen in Fig.~\ref{fig:Fe}. Since epitaxial Fe has a strong magneto-crystalline anisotropy the magnetization will in general not be aligned to the external magnetic field due to the anisotropy fields. The ISHE and SR effects are, however, only dependent on the angle of magnetization $\Theta_{\text{M}}$. For this reason, an additional rescaling of the angle axis is required. A numerical analysis has been performed where $\Theta_\text{M}$ has been calculated for the data measured at $\Theta_{\text{H}}$. For this $K_1/M_s$ ($K_1$: cubic anisotropy constant, $M_s$: saturation magnetization) has been extracted from the dependence of $H_\text{FMR}$ on the frequency (Kittel fit\cite{Kittel}). For RF-sputtered polycrystalline samples which are isotropic $\Theta_{\mathrm{H}}$ and $\Theta_{\mathrm{M}}$ are identical. However, in the case of epitaxial Fe they can differ more than $10^\circ$.

After the angle rescaling the angular dependent measurements of the Fe bilayers can be discussed: In these bilayers (Fig.~\ref{fig:Fe}) in-plane excited AHE seems to be equally prominent as in-plane excited AMR, as can be recognized from the lack of side-maxima and also from the voltage amplitudes of the fits shown in Table~\ref{tab:results}. This is a main difference to the Py case where AMR is strictly dominant. The AHE excited by the out-of-plane dynamic magnetic field is rather small (as it also is in the case for Py). The difference is not due to the difference in growth (poly- or single-crystalline) of the FM layers. For instance, recent results on also polycrystalline CoFeB/Pt and CoFeB/Ta layers show that there AHE is the only dominant effect while AMR is almost negligible\cite{Andres2017}. The weight of the different spin rectification contribution is reflecting only the strength of the different effects (AHE, AMR) in the ferromagnetic material. The different capping materials (insulator, respectively metal) is changing the relative contribution of the SR effects onto $V_{\mathrm{asym}}$. Additionally the epitaxial Fe samples, especially Fe/MgO, show characteristic features around angles, where the external field is oriented equidistant between the magnetic hard (e.g. 45$^\circ$) and easy axis (e.g. 0$^\circ$) of Fe. This is due to the intrinsic magnetic anisotropy influencing the angular dependencies of ISHE and SR. 

\begin{figure}
\includegraphics[width =0.95 \columnwidth]{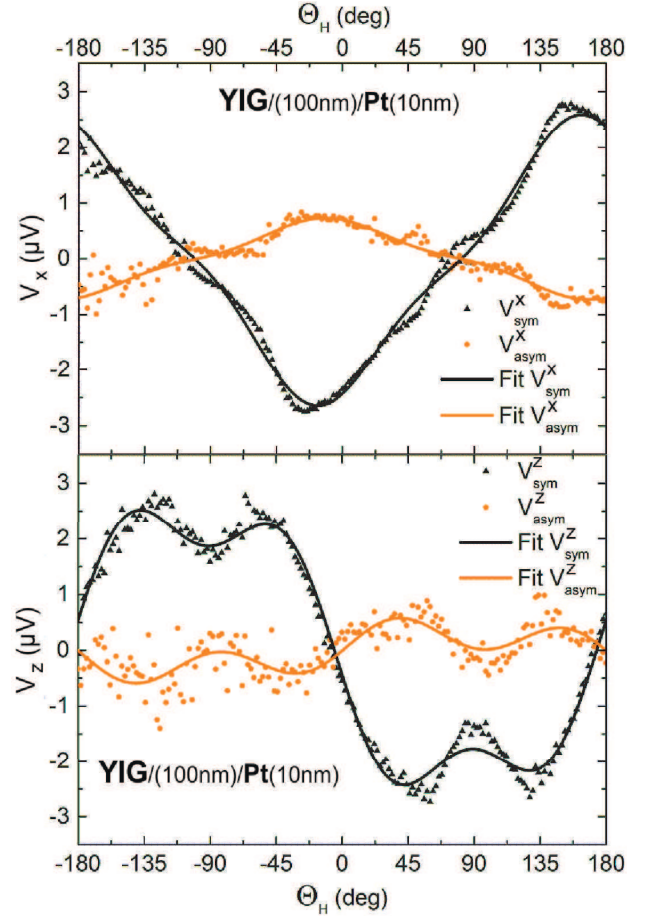}
\caption{\label{fig:YIGPt2} Angular dependent spin pumping measurements of YIG(100nm)/Pt(10nm) at 6.4 GHz excitation frequency with contacts transverse (top graph) and parallel (bottom graph) to the direction of the stripline antenna.}
\end{figure}

\begin{table*}
	\centering
		\begin{tabular} {ccrclrclrclcrclrclcrcl}
		$V_{\mathrm{sym}}/V_{\mathrm{asym}}$ & Contacts & \multicolumn{3}{c}{$V^{h_x}_{\mathrm{ISHE}}$ ($\mu V$)} & \multicolumn{3}{c}{$V^{h_y}_{\mathrm{ISHE}}$ ($\mu V$)} &\multicolumn{3}{c}{$V^{h_x}_{\mathrm{AHE}}$ ($\mu V$)} & ~ &\multicolumn{3}{c}{$V^{h_y}_{\mathrm{AHE}}$ ($\mu V$)}&\multicolumn{3}{c}{$V^{h_x,j_z}_{\mathrm{AMR}}$ ($\mu V$)} & ~ &\multicolumn{3}{c}{$V^{h_y,j_z}_{\mathrm{AMR}}$ ($\mu V$)}\\
			\hline
											
\multirow{2}{*}{$V_{\mathrm{sym}}$}	&	transverse & \multicolumn{3}{c}{amb.} &\multicolumn{3}{c}{{\bf -1.35}$^\star$} & \multicolumn{3}{c}{amb.} & ~ & -0.02 &\color[rgb]{0.5,0.5,0.5}$\pm$&\color[rgb]{0.5,0.5,0.5}0.01& \multicolumn{3}{c}{amb.} & ~ & -0.04 & \color[rgb]{0.5,0.5,0.5}$\pm$ & \color[rgb]{0.5,0.5,0.5}0.02 \\

																		&	parallel & \multicolumn{3}{c}{{\bf -1.48}$^\dagger$} & {\bf -1.82} & \color[rgb]{0.5,0.5,0.5}$\pm$&\color[rgb]{0.5,0.5,0.5}0.03& -0.19 & \color[rgb]{0.5,0.5,0.5}$\pm$&\color[rgb]{0.5,0.5,0.5}0.04 & ~ & 0.05 &\color[rgb]{0.5,0.5,0.5}$\pm$&\color[rgb]{0.5,0.5,0.5}0.02& \multicolumn{3}{c}{0$^\dagger$} & ~ & -0.09 & \color[rgb]{0.5,0.5,0.5}$\pm$&\color[rgb]{0.5,0.5,0.5}0.02\\
			\hline
											
\multirow{2}{*}{$V_{\mathrm{asym}}$}&	transverse &  \multicolumn{3}{c}{0}  & \multicolumn{3}{c}{0}& {\bf 0.53}&\color[rgb]{0.5,0.5,0.5}$\pm$&\color[rgb]{0.5,0.5,0.5}0.03 & ~ & 0.00 &\color[rgb]{0.5,0.5,0.5}$\pm$&\color[rgb]{0.5,0.5,0.5}0.01& 0.20&\color[rgb]{0.5,0.5,0.5}$\pm$&\color[rgb]{0.5,0.5,0.5}0.03 & ~ & 0.03 &\color[rgb]{0.5,0.5,0.5}$\pm$ & \color[rgb]{0.5,0.5,0.5}0.02\\
																		&	parallel & \multicolumn{3}{c}{0} & \multicolumn{3}{c}{0} & 0.11&\color[rgb]{0.5,0.5,0.5}$\pm$&\color[rgb]{0.5,0.5,0.5}0.03 & ~ & -0.01 &\color[rgb]{0.5,0.5,0.5}$\pm$&\color[rgb]{0.5,0.5,0.5}0.02& {\bf 0.65}&\color[rgb]{0.5,0.5,0.5}$\pm$&\color[rgb]{0.5,0.5,0.5}0.04 & ~ & -0.08&\color[rgb]{0.5,0.5,0.5}$\pm$&\color[rgb]{0.5,0.5,0.5}0.03 \\
			\hline
		\end{tabular}
	\caption{Results of the angular spin pumping measurements: symmetric and antisymmetric voltage amplitudes of YIG/Pt with contacts transverse and parallel to the microwave antenna. Values marked with * are ambiguous (see text). The voltage of the effects mainly contributing are marked in bold. The out-of-plane ISHE voltage with transverse contacts marked with $^\star$ has the same shape as the AHE and could easily be confused with it, but the comparison with the measurement with parallel contacts confirms it as an ISHE voltage. In-plane ISHE in the parallel contacts case marked with $^\dagger$ cannot be distinguished from in-plane AMR. In-plane AMR and AHE in the symmetrical voltage are estimated small since out-of-plane AMR and AHE are also small despite of relatively high out-of-plane excitation fields.}
	\label{tab:results2}
\end{table*}

\begin{figure}
\includegraphics[width =0.95 \columnwidth]{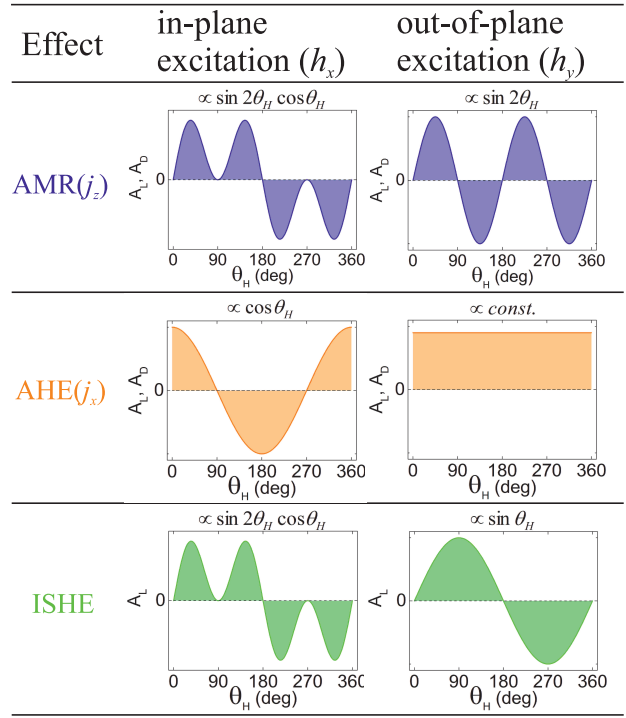}
\caption{\label{fig:SR6} Theoretical in-plane magnetization angular dependencies of spin rectification effects and ISHE with contacts parallel to the microwave antenna and different dynamic external magnetic field geometries, adapted from Harder et.~al. \cite{Harder2016}. $\Theta_{\mathrm{H}}$ is the magnetic field angle (defined in Fig.~\ref{fig:SRKS}), $A_{\text{L}}$ and $A_{\text{D}}$ are denoting the amplitudes of the effects contributing to the symmetric voltage (L: Lorentzian) and the antisymmetric voltage (D: Dispersive).}
\end{figure}

In addition, to compare with the measurements of the bilayers with metallic FM, a bilayer with an insulator magnetic material was measured with the same setup. For this YIG(100 nm)/Pt(10nm), where AMR and AHE are suppressed, was chosen and the results are shown in Fig.~\ref{fig:YIGPt2} (top graph) and Table~\ref{tab:results2}. A surprisingly non-vanishing antisymmetric voltage with angular dependencies similar to AMR and AHE can be seen. The symmetric voltage amplitude seems to be consisting mainly of an ISHE contribution and of a contribution $\propto \cos(\Theta_\mathrm{H})$ which can be either in-plane ISHE or out-of-plane AHE, as shown in Fig.~\ref{fig:SR5}. In order to understand this behavior we performed a second measurement with electrical contacts parallel to the stripline antenna ($z$-direction), measurements shown in Fig.~\ref{fig:YIGPt2} (bottom graph). In this contact geometry the ISHE and SR effects have different angular dependencies as shown in Fig.~\ref{fig:SR6}. Here in-plane excited ISHE and AMR have the same angular dependence, but the out-of-plane excited ISHE exhibits an unique $\cos(\Theta_\mathrm{H})$ dependence. To fit the measured data with contacts parallel to the antenna following equations has been used:

\begin{equation}
\begin{split}
V^{z}_{\mathrm{sym}} ~=~~ & V^{h_x}_{\mathrm{ISHE,AMR}}\sin(2\Theta_\mathrm{H})\cos(\Theta_\mathrm{H}) ~+\\
								 & V^{h_y}_{\mathrm{ISHE}}\sin(\Theta_\mathrm{H}) + V^{h_x}_{\mathrm{AHE}}\cos(\Theta_\mathrm{H}) ~+\\
								 & V^{h_y}_{\mathrm{AHE}} + V^{h_y,j_z}_{\mathrm{AMR}}\sin(2\Theta_\mathrm{H}).\\ & \\
V^{z}_{\mathrm{asym}} =~~ & V^{h_x}_{\mathrm{AHE}}\cos(\Theta_\mathrm{H}) +V^{h_y}_{\mathrm{AHE}} ~+\\
								 & V^{h_x}_{\mathrm{AMR}}\sin(2\Theta_\mathrm{H})\cos(\Theta_\mathrm{H}) ~+\\
								 & V^{h_y,j_z}_{\mathrm{AMR}}\sin(2\Theta_\mathrm{H}).
\end{split}
\label{fit2}
\end{equation}

Looking at Table~\ref{tab:results2} the out-of-plane ISHE contribution $V^{h_y}_{\mathrm{ISHE}}$ in transverse contacts ($-1.82 \mu V$) has almost the same value as the one of parallel contacts ($-1.35 \mu V$). The discrepancy is roughly reflecting the difference in sample width (2 mm) and length (3 mm) where the DC contacts have been applied to. The term $\propto \cos(\Theta_\mathrm{H})$ used for the fit of $V_\mathrm{sym}$ in Fig.~\ref{fig:YIGPt2} (top graph) is therefore confirmed as out-of-plane ISHE contribution.

To understand the enhancement of the out-of-plane magnetic microwave field component it is needed to consider that the size of the YIG/Pt sample is much smaller than the Fe and Py samples, its dimensions being closer to the width of the stripline antenna. This changes the distribution of the microwave fields and therefore enlarges the out-of-plane field components to the extent that they can be comparable in magnitude to the in-plane fields. In Fig.~\ref{fig:YIGPt2} (bottom graph) we also see a non-vanishing antisymmetric voltage with AMR-like dependence. Other authors also reported rectification effects in YIG/Pt bilayers in spin pumping experiments at room temperature caused by spin Hall magnetoresistance (SMR) \cite{Saitoh2014, Wu2016, Fang}. SMR is a rectification effect occuring in bilayers consisting of a FM insulator and a NM layer, where a spin current induced by spin Hall effect (SHE) forms a spin accumulation at the interface. When the magnetization is aligned parallel to the polarization of the accumulation, fewer spin currents can enter the FM layer and spin back-flow induces an additional charge current by ISHE reducing the resistivity of the NM layer. This in-plane angular dependent change in resistivity induces effects similar to AMR and AHE. Additionally the magnetic proximity effect can also contribute to spin rectification in YIG/Pt bilayers \cite{Caminale2016, Saitoh2013}: a ferromagnetic layer in contact to Pt can induce a finite magnetic moment in Pt near the interface because of the high paramagnetic susceptibility of Pt. This thin ferromagnetic Pt film can also exhibit spin rectification by itself with the same angular dependence. This is also true in metallic systems but there the spin rectification generated by the FM layer is dominating. Summarizing this section we have shown that the symmetric voltage of YIG/Pt is mainly consisting of ISHE contributions and the antisymmetric voltage is indeed small but not negligible and consisting of SMR induced rectification.

Furthermore, the results from the analysis of the YIG/Pt can be used to interpret the ISHE contribution in Py/Pt, seen in Fig.~\ref{fig:Py}: There the side-maxima of the symmetric amplitude are stronger pronounced than the ones of Py/Al. This is due to the reduction of the amplitude of the main-maximum of $V_{\mathrm{sym}}$ of Py/Pt. The reason for this is the opposite sign of the ISHE to the SR contributions. As shown in Table~\ref{tab:results2} ISHE from the YIG/Pt measurements shows a negative sign. The sign of the ISHE voltage is determined by the sign of spin Hall angle, the direction of the spin polarization and the direction of the spin current. They are all the same for both Py/Pt and YIG/Pt, therefore, the voltages generated by ISHE in Py/Pt and YIG/Pt should have the same sign.

In Tables~\ref{tab:results} and ~\ref{tab:results2} some values of the fits have not been shown, indicated by ``amb.". An intrinsic limitation of the analysis procedure is present when rotating the external magnetic field in-plane and investigating in-plane excited effects. According to Equation~\ref{verw} an ambiguity exists with the main in-plane contributions of this measurement configuration: ISHE, AMR and AHE are mathematically linearly dependent. Therefore, the absolute values obtained from the fits for the Py/Pt, Fe/Pt and YIG/Pt from $V_\mathrm{sym}$ may not be relevant. Nevertheless, the overall angular dependence and the $V_\mathrm{asym}$ data, where no ambiguity is present, support the interpretations shown in this paper.

\begin{equation}
\begin{split}
\cos(2\Theta_\mathrm{H})\cos(\Theta_\mathrm{H}) = & [2\cos^2(\Theta_\mathrm{H})-1]\cos(\Theta_\mathrm{H})\\ = & 2\cos^3(\Theta_\mathrm{H})-\cos(\Theta_\mathrm{H}).
\end{split}
\label{verw}
\end{equation} 

In summary, we have shown that the spin rectification effect does scale differently in Fe, Py and YIG bilayer systems, as summarized in Table~\ref{tab:results} and~\ref{tab:results2}: While AMR is more pronounced than AHE in RF magnetron sputtered Py, AHE seems to be equal in magnitude for epitaxial Fe systems. Spin rectification with an angular dependence similar to AMR is appearing in the antisymmetric Lorentzian shape in nanometer thin YIG/Pt bilayer films originating from the spin Hall magnetoresistance. The symmetric signal of YIG/Pt is mainly consisting of equal ISHE contributions excited by in- and out-of-plane dynamic magnetic fields. In epitaxial Fe systems the effects due to non-collinearity between the external field and the magnetization needs to be taken into account.

The Carl Zeiss Stiftung is gratefully acknowledged for financial support.

\bibliographystyle{apsrev4-1}

\end{document}